\documentclass[english,aps,preprint]{revtex4}
\usepackage[T1]{fontenc}
\usepackage[utf8]{luainputenc}
\usepackage{graphicx}

\makeatletter
\@ifundefined{textcolor}{}
{%
 \definecolor{BLACK}{gray}{0}
 \definecolor{WHITE}{gray}{1}
 \definecolor{RED}{rgb}{1,0,0}
 \definecolor{GREEN}{rgb}{0,1,0}
 \definecolor{BLUE}{rgb}{0,0,1}
 \definecolor{CYAN}{cmyk}{1,0,0,0}
 \definecolor{MAGENTA}{cmyk}{0,1,0,0}
 \definecolor{YELLOW}{cmyk}{0,0,1,0}
}

\makeatother

\usepackage{babel}
\begin{document}

\preprint{This line only printed with preprint option}

\title{Nonlinear Radiation Damping of Nuclear Spin Waves and Magnetoelastic
Waves in Antiferromagnets }

\author{Alexander V. Andrienko}

\email{direct22@front.ru}

\affiliation{National Research Centre “Kurchatov Institute”, 123182 Moscow, Russian
Federation}

\author{Vladimir L. Safonov}

\email{vlsafonov@magbiodyn.com}

\affiliation{Mag and Bio Dynamics, Inc., Arlington, TX 76001; Tarrant County College,
Fort Worth, TX 76119.}
\begin{abstract}
Parallel pumping of nuclear spin waves in antiferromagnetic CsMnF$_3$
at liquid helium temperatures and magnetoelastic waves in antiferromagnetic
FeBO$_3$ at liquid nitrogen temperature in a helical resonator was
studied. It was found that the absorbed microwave power is approximately
equal to the irradiated power from the sample and that the main restriction
mechanism of absortption in both cases is defined by the nonlinear
radiation damping predicted about two decades ago. We believe that
the nonlinear radiation damping is a common feature of parallel pumping
technique of all normal magnetic excitations and it can be detected
by purposeful experiments.
\end{abstract}
\maketitle
76.50.+g, 76.90.+d

\section{introduction}

Microwave parametric resonance of normal magnetic oscillations, such
as electronic spin waves, nuclear spin waves and magnetoelastic waves
is a powerful tool to study linear and nonlinear properties of magnetoordered
systems (ferromagnets, antiferromagnets and ferrites) \cite{suhl,schloemann,ZLS,L,GM,VS}
. Parallel pumping of magnetic excitations, in which the microwave
magnetic polarization is parallel to the external magnetic field,
is one of the most convenient and popular methods of parametric resonance.
A microwave magnetic field $\mathbf{h}(t)$ enhanced by a microwave
resonator is applied to the sample parallel to its equilibrium magnetization,
which is parallel to the steady external magnetic field $\mathbf{H}$.
The alternating magnetic field excites the parametric resonance of
the form $\omega_p = \omega_{\mathbf{k}} + \omega_{-\mathbf{k}}$,
where $\omega_p$ is the pumping field frequency and $\omega_{\mathbf{k}} = \omega_{-\mathbf{k}}$
are the half-pump frequencies of excited in the sample parametric
pair of waves with oppositly oriented wave vectors ${\mathbf{k}}$ 
and $-\mathbf{k}$. 

The excited waves amplitudes grow exponentially when the microwave
field amplitude $h$ exceeds the parametric resonance threshold $h_c$.
According to S theory \cite{ZLS,L}, this growth is restricted by
the nonlinearities of the magnetic system, which are exhibited by
a) phase mismatching of the forced magnetic oscillations with the
microwave field and b) by the positive nonlinear magnetic relaxation
due to nonlinearities of magnetic system. In this theoretical picture
the resonator cavity is considered to be just as an ancillary system
that enhances microwave field amplitude on a sample. No other effect
associated with the microwave resonator cavity is assumed in this
small sample approximation approach. 

Actually, the process of parallel pumping includes two steps. First,
the external microwave source excites the same frequency $\omega_p$
microwave magnetic oscillation of the resonator cavity. Second, this
magnetic oscillation is absorbed by the parametic pair ($\omega_{\mathbf{k}}$
and $\omega_{-\mathbf{k}}$) of magnetic excitations of the sample.
In principle, one can expect a backward radiation of the parametric
pairs and a bunch of associated effects in the system of two interacting
in resonance oscillations. However, in the simple picture of small
sample approximation this backward radiation is assumed to be negligibly
small compared to absorption; only an energy flow from the microwave
pump to the sample occurs. 

Notwithstading that examples of non-trivial role of microwave resonator
to the process of parallel pumping of magnetic excitations have already
been discussed in Refs.\cite{S jmmm,SY,ASirrad,AShumps,ASY}, these
facts did not attract much attention and the small sample approximation
approach is still in wide use for the description of parallel pumping
of magnetic oscillations. The main focus of the present paper is to
demonstrate that theoretically predicted two decades ago \cite{S jmmm,SY}
nonlinear radiation damping, effect due to backward irradiation of
parametric pairs to the resonator, is a common and dominant feature
in the process of parallel pumping of magnetic excitations. 

In this paper we studied parallel pumping in a helical resonator of
a) nuclear spin waves in an antiferromagnetic CsMnF$_3$ at liquid
helium temperatures and b) magnetoelastic waves in an antiferromagnetic
FeBO$_3$ at liquid nitrogen temperatures. 

The concept of nuclear spin waves was introduced by de Gennes et al
\cite{NSW}. The nuclear spin wave denotes the magnetic excitation
of mixed electronic and nuclear spin oscillations that is situated
in the nuclear magnetic resonance frequency range. The most remarkable
property of these excitations is that, at liquid helium temperatures,
they exhibit the coupled oscillations of two completely different
in their magnetic properties subsystems. The electronic spins are
ordered while the state of the nuclear spins is paramagnetic; the
polarization is no more than several percent. As a result of mixing
of these two subsystems by hyperfine interaction, the frequency of
the electronic spin waves increases, and the nuclear magnetic resonance
frequency $\omega_{n,0}$ decreases and becomes noticeably lower than
the Larmor precession frequency $\omega_{n}$ of nuclear spins. In
other words, there arises the so-called dynamic nuclear magnetic resonance
pulling and the band of nuclear spin waves $\omega_{n,k}$:

\begin{equation}
\omega_{n,k}=\omega_n \left[ 1- \left( \frac{\gamma H_{\Delta,hf}}{\omega_{e,k}}\right)^2 \right]^{1/2},                     \label{1}
\end{equation}where $\omega_{e,k}=\gamma [H(H+H_D) + H_{\Delta,hf}^2 + (\alpha k)^2 ]^{1/2}$
is the frequency of electronic spin wave, $H_D$ is the Dzyaloshinskii
field, $H_{\Delta,hf}^2 \propto 1/T$ is the gap due to hyperfine
interaction, $\alpha$ is the exchange constant and $\gamma$ is the
gyromagnetic ratio. The detailed review of nuclear spin wave properties
in weakly anisotropic antiferromagnets is given in Ref.\cite{AOSY}.

Magnetoelastic waves describe normal modes of linearly coupled elastic
waves and electronic spin waves in magnetoordered crystals. So far
as the magnetoelastic waves contain both elastic and magnetic components,
they can be excited both by elastic vibrations and by alternating
magnetic field. One of the most interesting objects to study magnetoelastic
waves is the high Néel temperature antiferromagnet FeBO$_3$ ($T_N = 348$
K). Parallel pumping of magnetoelastic waves in this crystal for the
first time was observed in Ref.\cite{AP}. The spectrum of magnetoelastic
waves in iron borate can be written as \cite{ASY}:

\begin{equation}
\omega_{me,k}=c_e k \left[ 1- \left( \frac{\gamma H_{\Delta,ef}}{\omega_{ek}}\right)^2 \right]^{1/2},                     \label{2}
\end{equation}where $c_e$ is the sound velocity, $H_{\Delta,ef}$ describes an
efficiency of linear interaction between spin and elastic subsystems,
$\omega_{e,k}=\gamma [H(H+H_D) + H_{\Delta,me}^2 + (\alpha k)^2 ]^{1/2}$
is the frequency of electronic spin wave and $H_{\Delta,me}$ is the
field which corresponds to magnetoelastic gap.

We show that beyond the small sample approximation the resonator oscillation
dynamics plays an extremely important role in the process of parametric
resonance of nuclear spin waves and magnetoelastic waves and gives
the dominant mechanism of parallel pumping restriction in both cases
by the nonlinear radiation damping.

\section{experiment}

The experimental absorbing cell is shown in Fig.1. The sample is placed
in an open helical resonator which is a half-wavelength dipole excited
by the pulsed microwave pumping field $\mathbf{h}(t)$. The inner
diameter of the helix equals 0.5 cm and the diameter of the copper
wire is 0.5 mm. To a first approximation the wire length needed to
make the helix is $\simeq \lambda / 2$ which is about 15 cm for 1
GHz. The effective volume of this resonator is estimated as $\sim 200$ ${\rm mm}^3$.
The effect of microwave absorption is detected by the receiving antenna.
This absorbing cell to study parallel pumping of nuclear spin waves
and magnetoelastic waves was used at different temperature conditions. 

\begin{figure}
\includegraphics[scale=1]{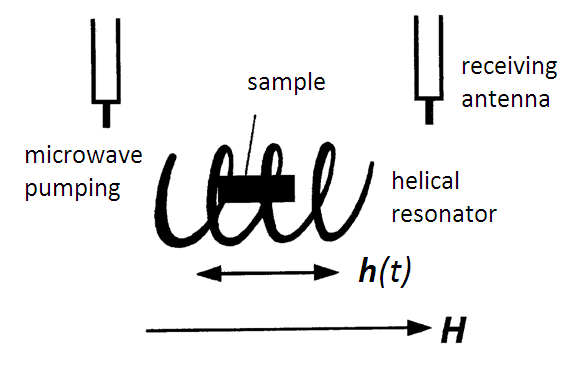}

\caption{Schematic diagramm of the experimental absorbing cell.}

\end{figure}

Parametric pairs of nuclear spin waves were excited by a pulsed ($300-2000$ $\mu {\rm s}$)
parallel microwave pump with repeating frequency $10-100$ Hz in the
helical resonator with the quality factor $Q \sim 300-500$ over a
wide range of frequencies $\omega_p = 600-1200$ MHz. The measurements
were made on single-crystal sample $v_s = 3 \times 3 \times 5$ ${\rm mm}^3$
of the easy-plane antiferromagnet CsMnF$_3$ ($T_N = 53.5$ K) at
liquid helium temperatures $T = 1.9 - 4.2$ K and magnetic fields
$H = 500 - 2000$ Oe. The ratio of the sample volume to the volume
of resonator was $v_s/v_R \sim 0.2$. The relaxation rate of parametically
excited spin waves estimated by the threshold amplituede was $\eta_{\mathbf{k}} / 2 \pi \sim 6-20$
kHz with the accuracy of 25 \%. 

A typical form of the microwave pump pulse passed thorough the resonator
is shown in the left side of Fig.2. There is a microwave absortpion
by the parametric pairs (upper part of the pulse) which is demonstrated
by the decrease of the pump pulse. At the end of the pulse one can
see a general phenomenon, a non-uniform time dependence with a peak
of the microwave radiation after the pump pulse, a typical oscillation
of the microwave pulse which appears above the threshold of parametric
resonance. We could observe this non-trivial radiation at $P/P_c - 1 \gg 1$.
The peak demonstrates a beating of magnetic oscillation of the resonator
cavity mode with the parametic pair \cite{ASirrad}. In this case
the lineshape of the cavity-sample system becomes splitted into two
humps \cite{AShumps} which is a direct indication that the small
sample approximation is not valid any more. Experimentally we observed
one peak if the pump frequency was equal to the frequency of the resonator
$\omega_p = \omega_R$ and up to three beating peaks if $\omega_p \ne \omega_R$
. It should be noted that below the threshold of parametric resonance
the microwave radiation after the pump pulse demonstrates just an
exponential decrease (see, curve 2 in Fig.2), which corresponds to
unloaded resonator cavity irradiation. 

\begin{figure}
\includegraphics[scale=1]{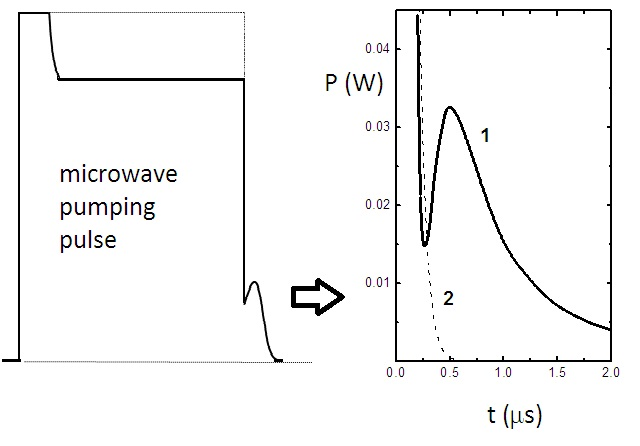}

\caption{\textbf{LEFT:} A typical form of the microwave pumping pulse passed
through the helical resonator. One can see a microwave absorption
by the sample (upper part) and a non-uniform radiation effect after
the end of mirowave pumping. \textbf{RIGHT:} Curve 1 demonstrates
a non-monotonic radiation power signal from the sample after the pump
pulse was turned off. The pumping power $P \approx 2000$ $P_c$. Curve
2 demonstrates the case when $P < P_c$, when just an exponentially
decreasing radiation from the resonator cavity is observed. The experimental
parameters are: $T=2.08$ K, $\omega_p/2\pi=1094$ MHz and $H=1840$
Oe. }

\end{figure}

\begin{figure}
\includegraphics[scale=1]{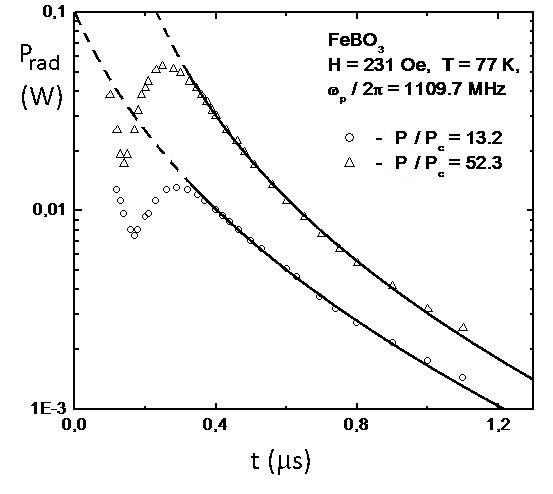}

\caption{Irradiation power of magnetoelastic waves versus time after the end
of the microwave pump pulse at two overcriticalities: $P/P_c = 13.2$
and$P/P_c = 52.3$ at $T = 77$ K, $\omega_p / 2\pi = 1109.7$ MHz
and $H = 231$ Oe. Solid lines describe theoretical fit (see the text). }
\end{figure}

Parametric pairs of magnetoelastic waves were excited in the $v_s \simeq 20$ ${\rm mm}^3$
sample of the ``easy-plane'' antiferromagnet FeBO$_3$ by the pulsed
microwave field of the frequency $\omega_p / 2\pi = 900-1200$ MHz
at magnetic fields $H = 30 - 500$ Oe at liquid nitrogen temperature
$T = 77$ K. The ratio of the sample volume to the volume of resonator
was $v_s/v_R \sim 0.1$. We observed similar effects of the non-uniform
radiation from the cavity-sample system after the end of mirowave
pump pulse as in the case of nuclear spin waves. Typical experimental
data of irradiation are shown in Fig.3.

We found very important feature of experiments with irradiation: the
radiation power is approximately equal to the absorption power. Thus,
the stationary state of parametric pairs is defined by the radiation
from the sample through the “sample-resonator” nonlinear interaction
.

\section{discussion}

Let us consider the monotonically decreasing time dependence of radiated
power behind the beating peak. The decrease of the parametric pairs
number $N_k(t)$ is described by the equation $ dN_k = -2\eta (N_k) dt $,
where $\eta (N_k)= \eta_k + \eta_{nl} N_k$ is the relaxation rate,
$\eta_k$ is the linear and $\eta_{nl} N_k$ is the nonlinear parts,
respectively. Integrating of this equation, one obtains

\begin{equation}
N_k(t) = \frac{\eta_k / \eta_{nl}}{u \exp [2\eta_k (t-t_0)] -1},                     \label{3}
\end{equation}where $u = 1 + \eta_k / \eta_{nl} N_k(t_0)$, $t_0$ is the starting
time ($t \ge t_0$).

If we assume that the nonlinear part of damping is entirely defined
by nonlinear radiation damping, then the radiated power $P_{rad}(t)$
can be expressed as

\begin{equation}
P_{rad}(t) = -\hbar \omega_p \frac{dN_k}{dt} \frac{\eta_{nl} N_k(t)}{\eta_k + \eta_{nl} N_k(t)} 
= \hbar\omega_p \frac{2\eta^2_k / \eta_{nl}}{ \{u \exp [2\eta_k (t-t_0)]-1 \}^2 }.          \label{4}
\end{equation}

\begin{figure}
\includegraphics[scale=1]{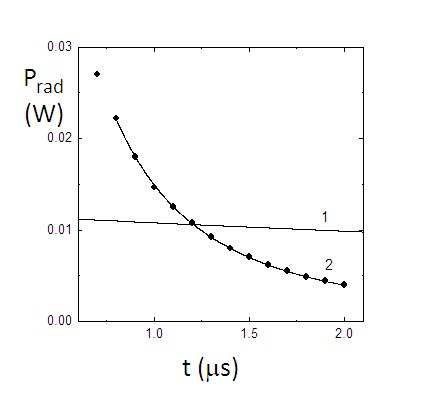}

\caption{Radiation power (dots) from the parametrically pumped nuclear spin
waves versus time in CsMnF$_3$ at $T = 2.08$ K, $\omega_p / 2\pi = 1094$
MHz and $H = 1840$ Oe. Curve 1 schematically demonstrate the radiation
power slope in the case of linear damping. Curve 2 demonstrate the
theoretical fit of formula (4) (see, the text). }
\end{figure}

\subsection{Nuclear Spin Waves}

A typical time slope for radiation power is shown in Fig. 4. Mean-square
fit using formula (4) with $t_0 = 0.8$ $\mu \rm{s}$ gives: $\hbar \omega_p \cdot 2\eta^2_k /\eta_{nl} = 1.8\cdot 10^{-4}$ W
and $\eta_k /\eta_{nl}N_k(0.8 \,\mu \rm{s})=9.05\cdot 10^{-2}$. The
linear relaxation rate calculated from the threshold of parallel pumping
is $\eta_k = 4.46\cdot 10^4$ s$^{-1}$. Thus we obtain $\eta_{nl} = 1.6\cdot 10^{-11}$
s$^{-1}$ and $\eta_{nl}N_k(0.8 \,\mu \rm{s}) = 4.93\cdot 10^5$ s$^{-1}$
which is one order greater than the linear relaxation rate $\eta_k$.
The number of parametric pairs at $t_0 = 0.8$ $\mu \rm{s}$ is equal
to $N_k(0.8 \,\mu \rm{s}) \simeq 3.1\cdot 10^{16}$. This esimate
for the number of parametric pairs is in agreement with the estimate
obtained in Ref.\cite{AndriN} from the susceptibility in the overthreshold
region. 

Note that the obtained result is stable to the variation of $\eta_k$.
For example, if we take linear relaxation rate, say, 40\% greater,
$\eta_k = 6.24\cdot 10^4$ s$^{-1}$, then from the fit one gets $\eta_{nl}N_k(0.8 \,\mu \rm{s}) = 4.64\cdot 10^5$
s$^{-1}$, $\eta_{nl} =1.4\cdot 10^{-11}$ s$^{-1}$ and $N_k(0.8 \,\mu \rm{s}) \simeq 3.3\cdot 10^{16}$.
We see that the accuracy of the threshold does not seriously affect
the nonlinear damping term due to relatively small value of linear
damping. 

Let us now compare experiment and theory. The theoretical formula
for the coefficient of nonlinear radiation damping can be expressed
in the form: 

\begin{equation}
\eta_{nl}^{(theor)} \simeq \xi_R \cdot 2 \pi \hbar Q \frac{V^2_k}{v_R},                     \label{5}
\end{equation}where $V_k$ is the coupling coefficient for the parametric pair with
the pump field in the resonator cavity, in other words, it is proportional
to an effective magnetic moment $\hbar {\partial \omega_{n,k}}/{\partial H} $
of excited wave. For nuclear spin waves one has \cite{AOSY,OzhYak}:

\begin{equation}
V_k = -\frac{1}{2} \frac{\partial \omega_{n,k}}{\partial H} = \frac{\omega_n^2}{4\omega_{n,k}} \frac{\gamma^4 (H_{\Delta,hf})^2 (2H+H_D)}{\omega_{e,k}^4}.                     \label{6}
\end{equation}

The factor $\xi_R$ in Eq.(5) depends on the geometry of resonator
cavity. For a rectangular resonator cavity one has $\xi_R=1$. For
a helical resonator a compression of half wavelength $\lambda/2$
to the length of helix $l$ occrus and it can result in $\xi_R \sim \lambda/2l$. 

Let us estimate theoretical nonlinear radiation damping, Eq.(5) for
the experiment shown in Fig.4, using the following parameters: $\omega_n=2\pi\cdot 666$ MHz,
$H_D=0$, $H^2_{\Delta,hf} = 6.4/T[K]$ kOe$^2$, $l \sim 1$ cm. One
gets: $\eta_{nl}^{(theor)} \sim 0.6\cdot 10^{-11}$ which of the order
of magnitude is in a good agreement with the obtained experimental
result.

\subsection{Magnetoelastic Waves}

Let us consider the experimental results shown in Fig.3 for magnetoelastic
waves. The linear relaxation rate calculated from the threshold of
parallel pumping in this case is $\eta_k = 3.2\cdot 10^5$ s$^{-1}$.
From the mean-square fit using formula (4) with $t_0 = 0.4$ $\mu \rm{s}$
one gets: 1) $\eta_{nl}N_k(0.4 \,\mu \rm{s})=0.55\cdot 10^6$ s$^{-1}$,
$N_k(0.4 \,\mu \rm{s}) \simeq 2.6\cdot 10^{16}$ for $P/P_c = 13.2$
and 2) $\eta_{nl}N_k(0.4 \,\mu \rm{s})=0.94\cdot 10^6$ s$^{-1}$,
$N_k(0.4 \,\mu \rm{s}) \simeq 4.4\cdot 10^{16}$ for $P/P_c = 52.3$.
For both cases we obtain the same experimental coefficient of nonlinear
radiation damping $\eta_{nl} = 2.1\cdot 10^{-11}$ s$^{-1}$. 

\begin{figure}
\includegraphics[scale=1]{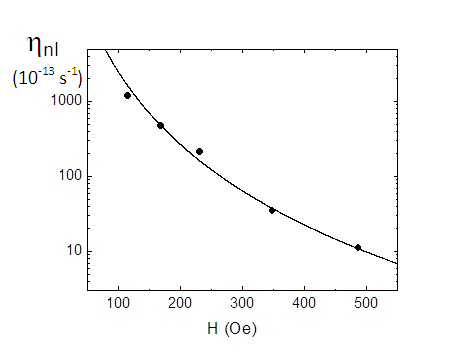}

\caption{Magnetic field dependence for the nonlinear radiation damping coefficient
of magnetoelastic waves in FeBO$_3$ at $T=77$ K and $\omega_p / 2 \pi = 1109.7$
MHz. Solid line is the theoretical fit. }

\end{figure}

In order to derive theoretical estimate, we find: \begin{equation}
V_k = -\frac{1}{2} \frac{\partial \omega_{me,k}}{\partial H} = \frac{(c_e k)^2}{4\omega_{me,k}} \frac{\gamma^4 (H_{\Delta,ef})^2 (2H+H_D)}{\omega_{e,k}^4}.                     \label{7}
\end{equation}

Thus, using Eq.(5) and the following parameters for the iron borate:
$c_e \simeq 4.8\cdot 10^5$ cm/s, $H_{\Delta,ef} \simeq 2$ kOe, $H_{\Delta,me}\simeq 2.2$ kOe,
$H_D \simeq 100$ kOe, $\alpha \simeq 0.08$ Oe$\cdot$cm, one gets:
$\eta_{nl}^{(theor)} \sim 2.9\cdot 10^{-11}$ s$^{-1}$, which of the
order of magnitude is in a good agreement with the obtained experimental
result. 

Dots in Fig. 5 show the magnetic field dependence of experimentally
obtained coefficient of nonlinear radiation damping. The solid line
represents the theoretical prediction of the field dependence. We
see a perfect fit within two orders of magnitude of experimental data
for nonlinear radiation damping.

\section{conclusion}

In this work we have experimentally confirmed that the nonlinear radiation
damping is the main mechanism of parametic instability restriction
during parallel microwave pumping of two different types of normal
magnetic oscillations, nuclear spin waves and magnetoelastic waves
in different antiferromagnets. The obtained results are in a good
agreement with the theory by the field and overthreshold dependencies
and are of the order of magnitude of the theoretical prediction. We
believe that the nonlinear radiation damping is a common feature of
parallel pumping technique and it can be detected by the purposeful
experiments with other types of normal magnetic oscillations in magnetoordered
systems. For example, a specific radiation after turning off the pump
of spin waves in YIG has already been observed in Ref.\cite{ZhitMel}
and it was not explained in the framework of small sample approximation.


\begin{thebibliography}{10}
\bibitem{suhl}H. Suhl, J. Phys. Chem. Solids \textbf{1}, 209 (1957).

\bibitem{schloemann}E. Schoemann, Phys. Rev. \textbf{116}, 827 (1959).

\bibitem{ZLS}V. E. Zakharov, V. S. L'vov, and S. S. Starobinets,
Usp. Fiz. Nauk \textbf{114}, 609 (1974) {[}Sov. Phys.- Usp. \textbf{17},
896 (1975){]}.

\bibitem{L}V. S. L’vov, Wave Turbulence under Parametric Excitation
(Springer-Verlag, Berlin, 1994).

\bibitem{GM}A. G. Gurevich and G. A. Melkov, Magnetization Oscillations
and Waves (CRC Press, Boca Raton, 1996).

\bibitem{VS}V. L. Safonov, Nonequilibrium Magnons (Wiley-VCH, Weinheim,
2013). 

\bibitem{S jmmm}V. L. Safonov, J. Magn. Magn. Mater. \textbf{97},
L1 (1991).

\bibitem{SY}V. L. Safonov and H. Yamazaki, J. Magn. Magn. Mater.
\textbf{161}, 275 (1996). 

\bibitem{ASirrad}A. V. Andrienko and V. L. Safonov, Pis'ma Zh. Eksp.
Teor. Fiz. \textbf{60}, 787 (1994) {[}JETP Lett. \textbf{60}, 800
(1994){]}.

\bibitem{AShumps}A. V. Andrienko and V. L. Safonov, Pis'ma Zh. Eksp.
Teor. Fiz. \textbf{6}2, 147 (1995) {[}JETP Lett. \textbf{6}2, 162
(1995){]}.

\bibitem{ASY}A. V. Andrienko, V. L. Safonov, and H. Yamazaki, J.
Phys. Soc. Jpn. \textbf{6}7, 2893 (1998). 

\bibitem{NSW}P. G. de Gennes, P. A. Pincus, F. Hartmann-Boutron,
and J. M. Winter, Phys. Rev. \textbf{129}, 1105 (1963).

\bibitem{AOSY}A. V. Andrienko, V. I. Ozhogin, V. L. Safonov, and
A. Yu. Yakubovskii, Usp. Fiz. Nauk, \textbf{161}, 1 (1991) {[}Sov.
Phys.- Usp. \textbf{34}, 843 (1991){]}.

\bibitem{AP}A. V. Andrienko and L. V. Podd'yakov, Zh. Eksp. Teor.
Fiz. \textbf{95}, 2117 (1989) {[}Sov. Phys.- JETP \textbf{68}, 1224
(1989){]}.

\bibitem{AndriN} A. V. Andrienko, Zh. Eksp. Teor. Fiz. \textbf{101},
1644 (1992) {[}Sov. Phys.- JETP \textbf{74}, 876 (1992){]}.

\bibitem{OzhYak} V. I. Ozhogin and A. Yu. Yakubovskii, Zh. Eksp.
Teor. Fiz. \textbf{67}, 287 (1974) {[}Sov. Phys.- JETP \textbf{40},
144 (1975){]}.

\bibitem{ZhitMel} V. S. Zhitnyuk and G. A. Melkov, Zh. Eksp. Teor.
Fiz. \textbf{75}, 1755 (1978) {[}Sov. Phys.- JETP \textbf{48}, 884
(1978){]}.\end{thebibliography}
\end{document}